\documentclass[twocolumn,superscriptaddress,amsmath,amssymb]{revtex4}
\usepackage[english]{babel}
\usepackage{graphicx}

\begin{document}

\title{Describing many-body bosonic waveguide scattering with the truncated Wigner method}

\author{Julien Dujardin}
\affiliation{D\'epartement de Physique, University of Liege, 4000 Li\`ege, Belgium}
\author{Thomas Engl}
\affiliation{Institut f\"ur Theoretische Physik, Universit\"at Regensburg, 93040 Regensburg, Germany}
\author{Juan Diego Urbina}
\affiliation{Institut f\"ur Theoretische Physik, Universit\"at Regensburg, 93040 Regensburg, Germany}
\author{Peter Schlagheck}
\affiliation{D\'epartement de Physique, University of Liege, 4000 Li\`ege, Belgium}

\begin{abstract}
We consider quasi-stationary scattering of interacting bosonic matter 
waves in one-dimensional waveguides, as they arise in guided atom lasers.
We show how the truncated Wigner (tW) method, which corresponds to
the semiclassical description of the bosonic many-body system on the
level of the diagonal approximation, can be utilized in order to
describe such many-body bosonic scattering processes.
Special emphasis is put on the discretization of space at the exact 
quantum level, in order to properly implement the semiclassical 
approximation and the tW method, as well as on the
discussion of the results to be obtained in the continuous limit.
\end{abstract}

\maketitle

\section{Introduction}

The perspective to realize atomtronic devices 
\cite{MicO04PRL,SeaO07PRA,PepO09PRL} 
as well as the exploration of transport features that are known from electronic 
mesoscopic systems \cite{BraO12S,BraO13S} have strongly stimulated the 
research on the dynamical properties of ultracold atoms in open systems.
While fermionic atoms provide direct analogies with the electronic case 
\cite{BraO12S,BraO13S,BruBel12PRA,KriO13PRL}, the use of a bosonic atomic 
species brings along new aspects and challenges for the atomic transport 
problem \cite{GutGefMir12PRB,Iva13EPJB}, related, in particular, with
the link between a mesoscopic Bose-Einstein condensate in a reservoir and
the microscopic dynamics of an ensemble of few interacting atoms
within a transistor-like device.
A particularly promising configuration for the experimental study of
these aspects is provided by the guided atom laser
\cite{GueO06PRL,CouO08EPL,GatO11PRL} in which atoms are
coherently outcoupled from a trapped Bose-Einstein condensate into an
optical waveguide.
A coherent atomic beam can thereby be created and injected onto engineered
optical scattering geometries.
This would allow one to study bosonic
many-body scattering at well-defined incident energy.

A theoretical modeling of such waveguide scattering processes within guided
atom lasers faces the challenge of dealing with an open system in a many-body
context, potentially involving a very large number of atoms in total.
Exact diagonalization methods therefore quickly encounter limitations when
describing such processes.
The Gross-Pitaevskii approximation, on the other hand, which has been used 
in a number of studies on guided atom laser processes 
\cite{LebPav01PRA,Car01PRA,PauRicSch05PRL,PauO05PRA,PauO07PRA}, is granted
to work in the mean-field limit of large atom densities and vanishing
atom-atom interaction strengths \cite{LieSeiYng00PRA}.
It is, however, known to break down in the presence of significantly 
strong interaction effects \cite{ErnPauSch10PRA}.

The truncated Wigner (tW) method \cite{GarZol,SteO98PRA,SinLobCas02JPB,Pol10AP}
appears to be a reasonable compromise in this context.
According to common understanding, this
method essentially consists in representing the time evolution
of a bosonic many-body quantum state by classical fields evolving
according to a Gross-Pitaevskii equation, whose initial values are
chosen such that they properly sample the initial quantum state under 
consideration.
They thereby provide a stochastic sampling of the approximated time
evolution of the many-body Wigner function defined in the phase space
of the bosonic field.
The tW method has been successfully applied in a number
of contexts involving dynamical processes of ultracold bosonic atoms.
This includes, most recently, scattering of atom laser beams
within one-dimen\-sional Bose-Hubbard systems \cite{DujArgSch15PRA}.
In this latter transport context, the tW method was shown to yield
quantitative predictions for the average transmitted current that
were shown to be in good agreement with complementary
matrix-product state calculations in the regime of moderate
on-site interaction strenghts \cite{DujArgSch15PRA}.

While it is possible to formulate the tW method in a 
functional manner \cite{OpaDru13JMP}, its practical implementation is 
most conveniently achieved by means of a discrete basis of the 
single-particle Hilbert space.
This is straightforward to accomplish for dynamical process that
are effectively taking place within spatially confined regions.
In that case, periodic boundary conditions can, e.g., be imposed
at a sufficiently large distance from the quantum many-body wave
packet to be studied, giving thereby rise to an effective 
discretization in momentum space.
This is, however, not a viable option for studying quasi-stationary
scattering processes which are generally characterized by an 
infinite spatial extension.
We therefore propose to discretize the \emph{position} space in
order to implement the tW method in this context,
as was effectively done in Ref.~\cite{DujArgSch15PRA}
A primary purpose of this paper is to introduce this discretization 
procedure in some detail and discuss its validity in the continuous
limit, both from an analytical and from a numerical point of view.

Moreover, we present in this paper an unconventional derivation 
of the tW method in the framework of the semiclassical 
van Vleck-Gutzwiller theory \cite{Gut}.
This allows us to identify the tW method with the
diagonal semiclassical approximation in the bosonic field-theoretical
context, and, as shown in Ref.~\cite{EngO14PRL}, to quantitatively
account for interference effects beyond tW. 
We note that the tW method can also be derived through other 
approaches such as
Wigner-Moyal expansions \cite{SteO98PRA,SinLobCas02JPB,Pol10AP}, 
quasiclasical corrections to the effective action \cite{Pol03PRA}, and the 
so-called semiclassical approximation in the context of the Keldysh approach 
\cite{Kam}. 
In those approaches, however, quantum corrections are perturbatively 
incorporated on top of a classical background given by the propagation of 
phase space distributions along classical trajectories that are
accounted for in an independent (i.e. incoherent) manner.
Interference effects involving different trajectories are essentially 
non-perturbative and require special resummation techniques 
that are justified only in the presence of small parameters, 
(typically the strength of interactions or the size of quantum fluctuations). 
While the van Vleck-Gutzwiller theory yields exactly the same (tW) 
approximation to leading (classical) order as the other approaches mentioned 
above, its key asset resides in the fact that it can readily incorporate 
quantum interference in a non-perturbative manner and describe, e.g.,
coherent backscattering in the Fock space of many-body systems \cite{EngO14PRL}.
This is our primary long-term motivation to advertise this approach in this 
article.
 
Our paper is organized as follows:
In Section \ref{sec:dis} we shall discuss in some detail the spatial
discretization procedure of one-dimensional waveguide scattering 
configurations.
Section \ref{sec:semi} is devoted to deriving the tW method in the framework
of the van Vleck-Gutzwiller theory.
We shall then argue in Section \ref{sec:noise} that the tW 
approach can be reformulated in terms of a stochastic Gross-Pitaevskii
equation whose noisy components exhibit well-defined statistical 
characteristics in the continuous limit.
These findings are confirmed by numerical results on transport
through a symmetric double barrier potential, as we show in Section 
\ref{sec:results}:
Both the average total current of atoms and its incoherent part tend to
finite values in the continuous limit.

\section{Discretization of space}

\label{sec:dis}

We consider a many-body scattering process of a coherent atomic
matter-wave beam within a waveguide.
This matter-wave beam is supposed to be created by a coherent
outcoupling process from a trapped Bose-Einstein condensate, as
is commonly done in guided atom lasers \cite{GueO06PRL,CouO08EPL,GatO11PRL}.
If we assume that only the transverse ground mode of the waveguide
is populated, we can describe this system by the many-body Hamiltonian
\begin{eqnarray}
  \hat{H} & = & \int_{-\infty}^{\infty} \hat{\psi}^\dagger(x)
  \left( - \frac{\hbar^2}{2m} \frac{\partial^2}{\partial x^2}
  + V(x) \right) \hat{\psi}(x) dx \nonumber \\
  & & + \frac{1}{2} \int_{-\infty}^{\infty} g(x) \hat{\psi}^\dagger(x)
  \hat{\psi}^\dagger(x)\hat{\psi}(x) \hat{\psi}(x) dx \nonumber \\
  & & + \int_{-\infty}^{\infty} \kappa(x) \left( \hat{\psi}^\dagger(x) \hat{b}
  + \hat{b}^\dagger \hat{\psi}(x) \right) dx + \mu \hat{b}^\dagger \hat{b}
  \label{eq:H}
\end{eqnarray}
where $\hat{\psi}^\dagger(x)$ and $\hat{\psi}(x)$ respectively represent 
the creation and annihilation operators of a bosonic particle at the
longitudinal position $x$ within the waveguide.
$m$ is the mass of the atoms, $V(x)$ describes a scattering potential 
(given, e.g., by a sequence of barriers) within the waveguide, and
$g(x)$ is the spatially dependent one-dimensional interaction strength
of the atoms in the waveguide.
It is approximately given by $g(x) \simeq 2 \hbar \omega_\perp(x) a_s(x)$
\cite{Ols98PRL} where $\omega_\perp$ is the transverse confinement frequency 
of the waveguide and $a_s$ is the (generally very small) s-wave scattering 
length of the atomic species under consideration. 
Both $\omega_\perp$ and $a_s$ can depend on the position in the waveguide, 
the former through a spatial variation of the transverse confinement, 
and the latter through a spatially dependent Feshbach tuning.

The trap is modeled by a single one-particle state with energy $\mu$ whose
associated creation an annihilation operators are given by $\hat{b}^\dagger$
and $\hat{b}$.
Trapped atoms are outcoupled to the waveguide through the spatially dependent 
(real) coupling strength $\kappa(x)$, which, e.g., in Ref.~\cite{GueO06PRL}
would model the effect of a radiofrequency field that flips the spin of the
atoms.
We shall in the following assume a macroscopically large initial population
$N\to \infty$ of the atoms in the trap in combination with a very small 
outcoupling strength $\kappa\to 0$, such that the trapped condensate is 
not appreciably affected by the outcoupling process on finite evolution times.
The time evolution of the system can then be effectively described by the 
equation
\begin{eqnarray}
  i\hbar \frac{\partial}{\partial t} \hat{\psi}(x,t) & = &
  \left( - \frac{\hbar^2}{2m} \frac{\partial^2}{\partial x^2}
  + V(x) - \mu \right) \hat{\psi}(x,t)  + \kappa(x) \hat{b} e^{-i\mu t/\hbar}
  \nonumber \\
  && + g(x) \hat{\psi}^\dagger(x,t)\hat{\psi}(x,t) \hat{\psi}(x,t)
\end{eqnarray}
for the time-dependent field operator $\hat{\psi}(x,t)$ describing atoms
in the waveguide.

In order to implement the tW method for this many-body
scattering problem, we first need to introduce a discretization procedure
of this spatially continuous quantum field equation.
As pointed out above, such a discretization cannot be defined in momentum
or energy space (e.g. through the introduction of periodic boundary 
conditions at $x=\pm L$ for some large $L\to\infty$) as this would not
be compatible with the formation of a quasi-stationary scattering state.
We therefore propose to discretize the \emph{position} space, namely through 
the introduction of a high-energy cutoff in momentum space at 
$p = \pm \pi \hbar / \delta$ for some effective spatial grid size 
$\delta\to 0$.
Correspondingly, we also modify the Hamiltonian of the free kinetic energy
close to the high-energy cutoff, such that it reads 
\begin{equation}
  \hat{H}_0 = E_\delta \int_{-\infty}^{\infty} \hat{\psi}^\dagger(x)
  \left( 1 - \cos(\delta \hat{p} / \hbar) \right) \hat{\psi}(x) dx
  \label{eq:H0}
\end{equation}
with $E_\delta = \hbar^2/(m \delta^2)$
and $\hat{p} = -i \hbar \partial/\partial x$.
This Hamiltonian is still diagonalized in the eigenbasis of the normalized
waves $\tilde{\phi}_k(x) = \exp(ikx)/\sqrt{2\pi}$ satisfying
$\int_{-\infty}^\infty \tilde{\phi}_k^*(x) \tilde{\phi}_{k'}(x) dx = \delta(k-k')$.
Their associated eigenvalues now read $E_k = E_\delta (1 - \cos k \delta )$
close to the cutoff while they are still approximately given by
$E_k \simeq \hbar^2 k^2 / (2m)$ for $k\delta\ll 1$.

In close analogy with the theory of spatially periodic systems,
we now introduce an effective Wannier basis through the spatially
localized functions
\begin{eqnarray}
 \phi_l(x) & = & \sqrt{\frac{\delta}{2\pi}} \int_{-\pi/\delta}^{\pi/\delta}
  \tilde{\phi}_k(x) e^{-i l k \delta} dk \\
  & = & \frac{1}{\sqrt{\delta}} \mathrm{sinc}
  \left(\frac{\pi}{\delta} (x - l \delta)\right)
\end{eqnarray}
for integer $l\in\mathbb{Z}$, with $\mathrm{sinc}(x) \equiv \sin(x)/x$, 
which can effectively be seen as Fourier series coefficients of the 
waves $\tilde{\phi}_k$.
They therefore satisfy 
$\int_{-\infty}^\infty \phi_l^*(x)\phi_{l'}(x) dx = \delta_{l l'}$
and thereby form an orthogonal basis set that spans the restricted
Hilbert space obtained after the introduction of the momentum cutoff.
Defining the corresponding creation and annihilation operators 
$\hat{\psi}_l^\dagger$, $\hat{\psi}_l$ such that we have
$\hat{\psi}(x) = \sum_{l=-\infty}^\infty \phi_l(x) \hat{\psi}_l$,
we can now reformulate the free kinetic Hamiltonian \eqref{eq:H0} such
that it reads
\begin{equation}
  \hat{H}_0 = \frac{1}{2} E_\delta \sum_{l=-\infty}^\infty \left( 
  2 \hat{\psi}_l^\dagger \hat{\psi}_l - \hat{\psi}_{l+1}^\dagger \hat{\psi}_l
  - \hat{\psi}_l^\dagger \hat{\psi}_{l+1} \right) \,.
\end{equation}
This expression is identical to the one that would be obtained from
a finite-difference approximation of the kinetic energy in the waveguide.

Altogether, we thereby obtain in the limit $\delta\to 0$
the effective Bose-Hubbard Hamiltonian
\begin{eqnarray}
  \hat{H} & = & \sum_{l=-\infty}^\infty \left[ (E_\delta - \mu + V_l) 
  \hat{\psi}_l^\dagger \hat{\psi}_l - \frac{E_\delta}{2} 
  \left( \hat{\psi}_{l+1}^\dagger \hat{\psi}_l
  + \hat{\psi}_l^\dagger \hat{\psi}_{l+1} \right) \right. \nonumber \\
  & & \left. + \frac{g_l}{2} \hat{\psi}_l^\dagger \hat{\psi}_l^\dagger
  \hat{\psi}_l\hat{\psi}_l + \kappa_l \left( \hat{\psi}_l^\dagger \hat{b}
  + \hat{b}^\dagger \hat{\psi}_l \right) \right] \label{eq:Hl}
\end{eqnarray}
where we introduce the definitions
$V_l \equiv V(l \delta)$, $g_l \equiv g( l \delta ) / \delta$, and
$\kappa_l \equiv \sqrt{\delta} \kappa(l \delta)$,
and where, for the sake of convenience, we redefine in Eq.~\eqref{eq:Hl} 
the zero of the energy scale such that it coincides with the energy of 
trapped atoms.
The time evolution of the discrete field operator $\hat{\psi}_l$ 
is then given by
\begin{eqnarray}
\label{eq:EOMdis}
  i\hbar \frac{\partial}{\partial t} \hat{\psi}_l(t) & = &
  (E_\delta - \mu + V_l) \hat{\psi}_l(t) - \frac{E_\delta}{2} 
  \left( \hat{\psi}_{l-1}(t) + \hat{\psi}_{l+1}(t) \right) \nonumber \\
  & & + g_l\hat{\psi}_l^\dagger(t) \hat{\psi}_l(t)\hat{\psi}_l(t)
  + \kappa_l \hat{b} \,.
\end{eqnarray}

\section{Semiclassical derivation of the truncated Wigner method}
\label{sec:semi}

Having accomplished the discretization of the exact quantum description, 
a semiclassical approach can be used in the regime of large particle 
numbers $N \gg 1$, which does not resort to the numerical solution of 
the operator-valued equations~(\ref{eq:EOMdis}). 
This approach has three levels of approximation. 
The first level is the purely classical limit consisting of the
propagation of the (discretized) Gross-Pitaevskii equation that 
is obtained from Eq.~(\ref{eq:EOMdis}) by the substitution 
$\hat{\psi}_l(t) \mapsto \psi_l(t)$ and 
$\hat{\psi}_l^\dagger(t)\mapsto \psi_l^*(t)$
where the latter classical fields are canonically conjugated 
variables. 
In a second step, a finite initial phase-space distribution
of the quantum many-body Wigner function is accounted for.
This phase-space distribution can contain non-classical correlations 
which is, however, not the case in our present problem.
Finally, a full-fledged semiclassical approximation is obtained by 
coherently adding amplitudes associated with classical trajectories,
in order to address quantum interference effects \cite{SimStr14PRA}. 
The second stage in this hierarchy of levels of approximation is the tW method.
Here we derive this method in an unconventional manner which is
based on a semiclassical approximation of the Feynman propagator 
of our effective Bose-Hubbard system \eqref{eq:Hl} in terms of
a coherent sum over solutions of the mean-field equations.

The notion of the term "semiclassical" in our approach is 
conceptually different from its meaning in the context of the 
Moyal-Wigner expansion used in quantum optics 
\cite{SteO98PRA,SinLobCas02JPB,Pol10AP}, the quantum corrections to the 
effective action of Ref.~\cite{Pol03PRA}, and the 
quasiclassical corrections within the Keldysh approach \cite{Kam}. 
In those approaches, the classical limit is identified 
in terms of the transport equations of \emph{probability} distributions 
in phase space (the so-called \textit{classical field} in the Keldysh 
language \cite{Kam}), while quantum corrections are systematically 
incorporated as perturbation series around this classical background. 
Although interference effects, with their charateristic non-perturbative 
dependence on $\hbar$, can be obtained within those frameworks by 
special resummation techniques, these techniques are in principle not 
suitable to tackle classically chaotic situations where all 
relevant physical forces governing the dynamics of the system under 
consideration are of the same order and no small parameter can be identified.

The semiclassical van Vleck-Gutzwiller approach, on the other hand, 
is particularly powerful precisely in this chaotic regime
where it allows one to predict universal interference effects.
Its key feature is that it approximates the quantum time evolution of the
system by a semiclassical superposition of \emph{amplitudes} and 
therefore explicitly accounts for quantum mechanical interference
through coherent double sums over classical paths. 
By working with amplitudes instead of probabilities, 
we can explicitely identify tW with the assumption that pairs of different 
classical solutions have uncorrelated actions.
Going beyond this so-called diagonal approximation and taking into
account systematic off-diagonal contributions,
the semiclassical approach has been used to predict interference effects 
beyond both tW and its quasiclassical corrections in good agreeement 
with numerical simulations \cite{EngO14PRL,EngUrbRic14xxx}.

Following Ref.~\cite{EngO14PRL}, we begin with the time evolution 
of an arbitrary many body state, 
$|\Psi(t)\rangle=\hat{U}(t)|\Psi(0)\rangle$,
governed by the time evolution operator 
$\hat{U}(t)=\exp(-i t \hat{H} / \hbar)$
associated with the Hamiltonian (\ref{eq:Hl}). 
Our Hilbert space
\begin{equation}
{\cal H}=\mathrm{span}\{|{\bold n}\rangle \equiv |\ldots, n_{-1},n_0, n_1,
\ldots\rangle, n_{l} \in \mathbb{N}_0 \}
\end{equation}
is spanned by the Fock states associated with the site basis,
where the discretized field operators act in the usual manner,
namely by raising or lowering the number of particles in a given site:
\begin{eqnarray}
\hat{\psi}_{l}^{\dagger}|\ldots,n_l,\ldots\rangle
&=&\sqrt{n_{l}+1}|\ldots,n_{l}+1,\ldots \rangle\,, \\
\hat{\psi}_{l}|\ldots,n_l,\ldots\rangle
&=&\sqrt{n_{l}}|\ldots,n_{l}-1,\ldots \rangle\,.
\end{eqnarray}

The starting point of our approach is to use of a different 
basis for ${\cal H}$, which is the formal equivalent of the position 
representation used in the usual derivation of the Feynman propagator 
in first-quantized systems. 
This new basis $|{\bold q}\rangle$ is given by the common eigenstates 
of the commuting set of hermitian operators
$\hat{q}_{l}=\sqrt{\hbar/2}(\hat{\psi}_{l}+\hat{\psi}_{l}^{\dagger})$
which are known in quantum optics as quadratures. 
Using the defining property
$\hat{q}_{l}|{\bold q}\rangle=q_{l} |{\bold q}\rangle$,
the orthonormality and completness relations
$\langle {\bold q}|{\bold q}'\rangle=\delta({\bold q}-{\bold q}')$ and
$\int d{\bold q}|{\bold q}\rangle \langle{\bold q}|=\hat{1}$
of the quadrature basis follow directly.

Having at hand a complete basis with continuous states, a path integral 
representation of the time evolution operator can be obtained in the usual way.
Contrary to the usual kinetic-plus-potential Hamiltonian in
the first-quantized case, the use of the eigenstates 
$|{\bold p}\rangle$ of the momentum quadratures
$\hat{p}_{l}=-i\sqrt{\hbar/2}(\hat{\psi}_{l}-\hat{\psi}_{l}^{\dagger})$
is not only useful but essential to get the amplitude 
$K({\bold q}^{\rm f},{\bold q}^{\rm i},t) \equiv 
\langle {\bold q}^{\rm f}|\hat{U}(t)|{\bold q}^{\rm i}\rangle$
of the propagation from an initial to a final quadrature state
\begin{equation}
\label{eq:K}
K({\bold q}^{\rm f},{\bold q}^{\rm i},t)= \int{\cal D}[{\bold q}(s),{\bold p}(s)]
{\rm e}^{i R[{\bold q}(s),{\bold p}(s)] / \hbar}
\end{equation}
in the form of an integral over paths ${\bold q}(s),{\bold p}(s)$ satisfying 
the boundary (shooting) conditions ${\bold q}(s=0)={\bold q}^{\rm i}$ and 
${\bold q}(s=t)={\bold q}^{\rm f}$. 
Here, the action functional is given in its Hamiltonian form,
\begin{equation}
\label{eq:R}
R[{\bold q}(s),{\bold p}(s)]=\int_{0}^{t} ds[\dot{{\bold q}}(s) 
\cdot {\bold p}(s)-H^{\rm cl}({\bold q}(s),{\bold p}(s))]
\end{equation}
where the classical Hamiltonian $H^{\rm cl}$ is obtained from 
the quantum one by the substitutions 
$\hat{\psi}_l\mapsto\psi_l$ and $\hat{\psi}_l^\dagger\mapsto\psi_l^*$
with $\psi_l\equiv (q_l + i p_l)/\sqrt{2\hbar}$.
Properly taking care of the Weyl ordering of operators, which yields 
the replacement $\hat{\psi}_l\hat{\psi}_l^\dagger+
\hat{\psi}_l^\dagger\hat{\psi}_l\mapsto 2\psi_l^*\psi_l$, 
we then obtain from Eq.~\eqref{eq:Hl} the classical Gross-Pitaevskii-type
Hamiltonian
\begin{eqnarray}
  H^{\rm cl} & = & \sum_{l=-\infty}^\infty \left[ 
      (E_\delta - \mu + V_l) \psi_l^* \psi_l - \frac{E_\delta}{2} 
  \left( \psi_{l+1}^* \psi_l + \psi_l^* \psi_{l+1} \right) \right. \nonumber \\
  & & \left. + \frac{g_l}{2} \psi_l^* \psi_l ( \psi_l^* \psi_l - 2  )
  + \kappa_l \left( \psi_l^* b + b^* \psi_l \right) \right]
    \label{eq:Hcl}
\end{eqnarray}
up to constant terms that are not important.

We now continue with a stationary phase analysis of the path integral, 
following Gutzwiller's pionering work in the 60's and 70's \cite{Gut}. 
This programme was accomplished in Ref.~\cite{EngO14PRL}.
It yields as final result the propagator
\begin{equation}
\label{eq:Ksc}
K({\bold q}^{\rm f},{\bold q}^{\rm i},t)\simeq \sum_{\gamma}\left|{\rm det}
\frac{\partial^{2}}{\partial{\bold q}^{\rm i} \partial{\bold q}^{\rm f}}
\frac{R_{\gamma}({\bold q}^{\rm f},{\bold q}^{\rm i},t)}{2\pi \hbar}
\right|^{1/2} {\rm e}^{iR_{\gamma}({\bold q}^{\rm f},{\bold q}^{\rm i},t)/\hbar}
\end{equation}
as a sum over all solutions of the classical shooting problem
(up to an extra phase $i\mu_{\gamma}\pi/4$ involving the Maslov
index $\mu_{\gamma}$, which is not important in the following).

The time evolution of the expectation value $\langle \hat{A} \rangle_{t} 
\equiv \langle \Psi(t)|A(\hat{\bold q},\hat{\bold p})|\Psi(t)\rangle$
of a properly (Weyl) ordered operator 
$\hat{A}=A(\hat{\bold q},\hat{\bold p})$ is given by
\begin{eqnarray}
\label{eq:Expect}
\langle \hat{A} \rangle_{t} &=&\int d{\bold q}^{\rm f}d{\bold q}^{\rm i} 
d{\bold q'}^{\rm i} \langle{\bold q'}^{\rm i}|\Psi(0)\rangle\langle 
\Psi(0)|{\bold q}^{\rm i}\rangle \nonumber \\ 
&\times& K({\bold q}^{\rm f},{\bold q}^{\rm i},t)^{*}
A({\bold q}^{\rm f},-i\hbar \partial / \partial {\bold q}^{\rm f})
K({\bold q}^{\rm f},{\bold q'}^{\rm i},t).
\end{eqnarray} 
Our goal is to obtain a semiclassical approximation for this expression.
As first step, we substitute the semiclassical propagator (\ref{eq:Ksc}) 
in Eq.~(\ref{eq:Expect}). 
At this level of approximation, we only assume that a typical action 
in Eq.~(\ref{eq:R}) is significantly larger than $\hbar$.
In that limit, for any smooth functions $F(x),G(x),H(x)$ we have
\begin{equation} 
F\left(-i\hbar\frac{\partial}{\partial {\bold q}}\right)
G({\bold q}){\rm e}^{\frac{i H({\bold q})}{\hbar}}=
F\left(\frac{\partial H({\bold q})}{\partial {\bold q}} \right)
G({\bold q}){\rm e}^{\frac{i H({\bold q})}{\hbar}}+{\cal O}(\hbar).
\end{equation}
Using $\partial R_{\gamma}({\bold q}^{\rm f},{\bold q'}^{\rm i},t) 
/ \partial {\bold q}^{\rm f} = {\bold p}^{\rm f}_{\gamma} 
({\bold q}^{\rm f},{\bold q}^{\rm i},t)$,
we obtain then
\begin{eqnarray}
&&A({\bold q}^{\rm f},-i\hbar \partial / \partial {\bold q}^{\rm f})
K({\bold q}^{\rm f},{\bold q'}^{\rm i},t) \nonumber \\ 
&\simeq& \sum_{\gamma}A({\bold q}^{\rm f},{\bold p}_{\gamma}^{\rm f}) \left|{\rm det}
\frac{\partial^{2}}{\partial{\bold q}^{\rm i} \partial{\bold q}^{\rm f}}
\frac{R_{\gamma}({\bold q}^{\rm f},{\bold q}^{\rm i},t)}{2\pi \hbar}
\right|^{1/2} {\rm e}^{i R_{\gamma}({\bold q}^{\rm f},{\bold q}^{\rm i},t)/\hbar}.
\end{eqnarray}
This then yields a double sum over classical trajectories in 
Eq.~\eqref{eq:Expect}. 
It is remarkably accurate even for moderate values of 
the classical actions and for arbitrary large times, and it sucessfully 
describes interference phenomena through the coherent sum over the 
oscillatory amplitudes that are associated with each trajectory.

In a further approximation, one makes use of the fact that the double sum 
over paths contains a large number of terms that effectively cancel 
each other in the presence of an average over initial and final positions 
as in Eq.~\eqref{eq:Expect}. 
Only pairs of trajectories with similar actions yield nonvanishing 
contributions due to phase cancellations, which gives rise to 
incoherent sums of slowly oscillatory terms with non-zero average. 
This is the essence of the \emph{diagonal approximation}, the 
standard tool to obtain leading, classical-like, contributions 
from coherent double sums representing semiclassical amplitudes. 
In the present context, the diagonal approximation is implemented by 
pairing $\gamma({\bold q}^{\rm f},{\bold q}^{\rm i},t)$ and 
$\gamma({\bold q}^{\rm f},{\bold q'}^{\rm i},t)$ for which 
${\bold q}^{\rm i}\simeq {\bold q'}^{\rm i}$. 
In the semiclassical limit, all smooth terms in the propagator are then 
evaluated at ${\bold q} =({\bold q}^{\rm i}+{\bold q'}^{\rm i})/2$, 
while the actions are expanded up to first order in 
${\bold q}^{\rm i}-{\bold q'}^{\rm i}$ according to
$R_{\gamma}({\bold q}^{\rm f},{\bold q}^{\rm i},t)-
R_{\gamma}({\bold q}^{\rm f},{\bold q'}^{\rm i},t) \simeq
- {\bold p}_{\gamma}^{\rm i}({\bold q}^{\rm f},{\bold q},t) 
\cdot ({\bold q}^{\rm i}-{\bold q'}^{\rm i})$.

The double sums are then transformed into a single sum 
yielding the diagonal approximation for $\langle \hat{A} \rangle_{t}$ as
\begin{eqnarray}
\langle \hat{A} \rangle_{t}^{\rm d}&=&
\int d{\bold q}^{\rm f}d{\bold q}^{\rm i} d{\bold q'}^{\rm i}
\langle{\bold q'}^{\rm i}|\Psi(0)\rangle \langle \Psi(0)|{\bold q}^{\rm i}\rangle 
\sum_{\gamma}A[{\bold q}^{\rm f},{\bold p}_{\gamma}^{\rm f}
({\bold q}^{\rm f},{\bold q},t)] \nonumber \\ 
&\times& \left|{\rm det}
\frac{\partial^{2}}{\partial{\bold q} \partial{\bold q}^{\rm f}}
\frac{R_{\gamma}({\bold q}^{\rm f},{\bold q},t)}{2\pi \hbar} \right|
{\rm e}^{-i{\bold p}_{\gamma}^{\rm i}({\bold q}^{\rm f},{\bold q},t) 
\cdot ({\bold q}^{\rm i}-{\bold q'}^{\rm i})/\hbar} .
\end{eqnarray}
This expression can be further simplified by noticing that 
$ \partial^{2} R_{\gamma}({\bold q}^{\rm f},{\bold q},t) /
\partial {\bold q} \partial {\bold q}^{\rm f} = 
- \partial {\bold p}^{\rm i}_{\gamma}({\bold q}^{\rm f},{\bold q},t)
/ \partial {\bold q}^{\rm f}$
is the Jacobian matrix of the transformation 
${\bold q}^{\rm f} \to {\bold p}^{\rm i}$, as these variables are 
functionally dependent through the solutions 
${\bold q}^{\rm f}={\bold q}^{\rm f}({\bold q},{\bold p}^{\rm i},t)$,
${\bold p}^{\rm f}={\bold p}^{\rm f}({\bold q},{\bold p}^{\rm i},t)$
of the classical equations of motion.
We can then change the integration over final coordinates by an 
integration over initial momenta. 
Together with the substitution
$({\bold q}^{\rm i},{\bold q'}^{\rm i}) \mapsto ({\bold q}, {\bold x})$
with ${\bold q} =({\bold q}^{\rm i}+{\bold q'}^{\rm i})/2$ and
${\bold x}={\bold q}^{\rm i}-{\bold q'}^{\rm i}$ (and with the replacement
${\bold p}^{\rm i} \equiv {\bold p}$), this finally yields 
\begin{equation}
\langle \hat{A} \rangle_{t}^{\rm d}=
\int d{\bold q}d{\bold p}A[{\bold q}^{\rm f}
({\bold q},{\bold p},t),{\bold p}^{\rm f}({\bold q},{\bold p},t)]
W_{\Psi(0)}({\bold q},{\bold p}), \label{eq:AtW}
\end{equation}
where we introduce the Wigner function associated with the state 
$|\Phi\rangle$ through
\begin{equation}
W_{\Phi}({\bold q},{\bold p}) =
\left(\prod_l \int \frac{d x_l}{2\pi\hbar}\right)
{\rm e}^{-i{\bold p} \cdot {\bold x} / \hbar}
\langle{\bold q}+ {\bold x}/2|\Phi\rangle 
\langle \Phi|{\bold q}- {\bold x}/2\rangle
\end{equation}
with the product $\prod_l$ going over all single-particle sites 
\cite{rem_exact}.
It is possible to reformulate Eq.~\eqref{eq:AtW} as
\begin{equation}
\langle \hat{A} \rangle_{t}^{\rm d}=
\int d{\bold q}d{\bold p}A({\bold q}, {\bold p})
W_{\Psi(0)}({\bold q},{\bold p},t)
\end{equation}
where $W_{\Psi(0)}({\bold q},{\bold p},t)$ is the time-dependent
Wigner function which evolves according to the well-known 
truncated Wigner equation.
This demonstrates that the tW method is obtained by neglecting 
off-diagonal terms in the semiclassical calculation of time-dependent 
expectation values of quantum observables.

For the waveguide scattering configuration under consideration,
the classical field equations to be numerically integrated in the
framework of the tW method are derived from Eq.~\eqref{eq:Hcl} and read
\begin{eqnarray}
  i\hbar \frac{\partial}{\partial t} \psi_l(t) & = &
  (E_\delta - \mu + V_l) \psi_l(t) - \frac{E_\delta}{2} 
  \left( \psi_{l-1}(t) + \psi_{l+1}(t) \right) \nonumber \\
  & & + g_l\left( |\psi_l(t)|^2 - 1 \right) \psi_l(t) + \sqrt{N} \kappa_l \,.
  \label{eq:TW}
\end{eqnarray}
Here we represent, in the limit of a large number of atoms $N\to\infty$, 
the Bose-Einstein condensate in the trap by a coherent state with the 
amplitude $\sqrt{N}$ and with vanishing relative uncertainty in its 
amplitude and phase.
The initial quantum state in the waveguide, on the other hand, is a vacuum
state that has to be correctly sampled in the framework of the truncated 
Wigner method.
More specifically, the initial field amplitudes in the discretized waveguide
have to be chosen as $\psi_l(0) = (A_l + i B_l)/2$ where $A_l$ and $B_l$ are
independent Gaussian random variables with zero mean and unit variance.
As a consequence, each site $l$ of the grid representing the empty waveguide
has initially the average classical density 
\begin{equation}
  \overline{|\psi_l(0)|^2} = 1/2 \label{eq:fl0}
\end{equation}
where the overline denotes the tW average over the random 
variables.
This pseudo-density of half a particle per grid point is to be subtracted
when calculating the average atomic density 
$\rho(x,t) = \langle \hat{\psi}^\dagger(x,t) \hat{\psi}(x,t)\rangle$
using the tW method:
this average density is evaluated as
\begin{equation}
  \rho(x=l\delta,t) = \frac{1}{\delta}
  \left( \overline{|\psi_l(t)|^2} - 1/2 \right) \,.
\end{equation}
The current $j(x,t) = \mathrm{Re}[ \langle \hat{\psi}^\dagger(x,t) \hat{p} 
\hat{\psi}(x,t)\rangle] / m$, on the other hand, is not articifially 
affected by the vacuum fluctuations; 
it is evaluated as
\begin{equation}
  j(x=l\delta,t) = \frac{\hbar}{2i m \delta^2} \left( 
  \overline{\psi_l^*(t)\psi_{l+1}(t) - \psi_{l+1}^*(t)\psi_l(t)} \right) \,.
  \label{eq:j}
\end{equation}

For both the density and the current we can define coherent components 
according to
\begin{eqnarray}
  \rho^{\rm coh}(l\delta,t) & = & \frac{1}{\delta} \left| \overline{\psi_l(t)}
  \right|^2 \,, \\
  j^{\rm coh}(l\delta,t) & = & \frac{\hbar}{2i m \delta^2} \left(
  \overline{\psi_l^*(t)} \, \overline{\psi_{l+1}(t)} - 
  \overline{\psi_{l+1}^*(t)}\,\overline{\psi_l(t)} \right) \,,
\end{eqnarray}
which result from a direct tW average of the complex
amplitudes $\psi_l$.
They would correspond to the mean-field predictions obtained by the 
classical Gross-Pitaevskii approach in the limit of large densities 
and vanishing atom-atom interaction strength.
The associated incoherent components are defined through
\begin{eqnarray}
  \rho^{\rm incoh}(l\delta,t) & = & \rho(l\delta,t) - 
  \rho^{\rm coh}(l\delta,t) \,, \label{eq:dinc} \\
  j^{\rm incoh}(l\delta,t) & = & j(l\delta,t) - j^{\rm coh}(l\delta,t) \,.
  \label{eq:jinc}
\end{eqnarray}
They approximately characterize the amount of quantum depletion in the
waveguide.

\section{The continuous limit}

\label{sec:cont}

\subsection{Quantum noise}

\label{sec:noise}

The tW method is certified to be valid in the semiclassical 
regime where the average population per single-particle state is large 
compared to unity.
However, this condition is no longer met in the continuous limit of
vanishing grid spacing $\delta\to 0$.
Indeed, assuming that the scattering process under consideration
gives rise to a well-defined average atom density $\rho(x,t)$ in
the waveguide, the corresponding discrete mean-field amplitudes would
essentially be given by $\psi_l(t) \simeq \sqrt{\delta \rho(l\delta,t)}$.
In the limit $\delta\to 0$ they would thereby become negligibly small as
compared to the vacuum fluctuations which still feature half a 
(pseudo-)particle per grid point according to Eq.~\eqref{eq:fl0}.

The purpose of this section is not to discuss the validity of the 
tW method in the continuous limit (which would obviously
require the comparison of numerical results with a complementary, and
genuinely quantum, method, such as exact diagonalization) 
but to argue that it yields consistent results in this limit.
We shall, for this purpose, introduce a Bogoliubov decomposition of the
tW amplitude according to $\psi_l(t) = \chi_l(t) + \varphi_l(t)$
where $\chi_l$ results from the unperturbed time evolution of the vacuum
fluctuation according to
\begin{eqnarray}
  i\hbar \frac{\partial}{\partial t} \chi_l(t) & = &
  (E_\delta - \mu + V_l) \chi_l(t) - \frac{E_\delta}{2} 
  \left( \chi_{l-1}(t) + \chi_{l+1}(t) \right) \nonumber \\
  & & + g_l\left( |\chi_l(t)|^2 - 1 \right) \chi_l(t) \label{eq:chi}
\end{eqnarray}
with the initial condition $\chi_l(0) = \psi_l(0)$.
This equation will not change the characteristics of the vacuum fluctuations
in the course of time evolution as it obviously corresponds to the truncated
Wigner modeling of an empty waveguide.
Through Eqs.~\eqref{eq:TW} and \eqref{eq:chi}, 
$\varphi_l\equiv\psi_l-\chi_l$ then evolves according to
\begin{eqnarray}
  i\hbar \frac{\partial}{\partial t} \varphi_l(t) & = &
  (E_\delta - \mu + V_l) \varphi_l(t) - \frac{E_\delta}{2} 
  \left( \varphi_{l-1}(t) + \varphi_{l+1}(t) \right) \nonumber \\
  && + g_l |\varphi_l(t)|^2 \varphi_l(t) + \sqrt{N} \kappa_l \nonumber \\
  && + g_l \left[ ( 2 |\chi_l(t)|^2 - 1 ) \varphi_l(t) + 
    \chi_l^2(t) \varphi_l^*(t) \right] \nonumber \\
  && + g_l \left[ ( 2 \chi_l(t) |\varphi_l(t)|^2 + \chi_l^*(t) \varphi_l^2(t) 
    \right] \,, \label{eq:varphi}
\end{eqnarray}
which effectively corresponds to a stochastic Gross-Pita\-evskii equation.
Indeed, the first two lines of Eq.~\eqref{eq:varphi} describe the mean-field
Gross-Pitaveskii modeling of the waveguide scattering process, with
$\varphi_l(0) = 0$ as initial condition in the waveguide.
The other two lines contain quantum noise terms that randomly perturb the
evolution of the coherent field $\varphi_l$ emitted by the outcoupling
process from the trap.

It is instructive to quantitatively evaluate the characteristics of 
this quantum noise in the continuous limit $\delta\to 0$.
In that limit, we can neglect the last line of Eq.~\eqref{eq:varphi}
since this particular noise term is suppressed to the noise terms in the
third line of Eq.~\eqref{eq:varphi} by a scaling factor $\sqrt{\delta}$
(noting that $g_l \sim 1/\delta$ and $\varphi_l \sim \sqrt{\delta}$
for $\delta\to 0$).
We can, furthermore, analytically integrate Eq.~\eqref{eq:chi}
by neglecting the effect of $\mu$, $V_l$ and $g_l$ as compared to $E_\delta$
(and by using the fact that noise amplitudes $\chi_l$ on adjacent
grid points are uncorrelated, in striking contrast to the coherent
amplitudes $\varphi_l$).
This yields
\begin{equation}
  \chi_l(t) = \sum_{l'=-\infty}^\infty \psi_l(0) i^{l-l'} 
  J_{l-l'}(E_\delta t/\hbar) e^{-i E_\delta t / \hbar}
\end{equation}
where $J_l$ denotes the Bessel function of the first kind of the order $l$.
We thereby obtain the temporal correlation function of the quantum noise as
\begin{equation}
  \overline{\chi_l^*(t)\chi_{l'}(t')} = \frac{1}{2} i^{l'-l} 
  J_{l'-l}\left(E_\delta (t'-t)/\hbar\right) e^{-i E_\delta (t'-t) / \hbar} \,,
\end{equation}
which reproduces Eq.~\eqref{eq:fl0} as special case for $t=t'$.

The third line of Eq.~\eqref{eq:varphi} can then be rewritten as 
$\xi_l(t) \varphi_l(t) + \eta_l(t) \varphi_l^*(t)$ with the effective
(real and complex) noise terms
\begin{eqnarray}
  \xi_l(t) & = & g_l ( 2 |\chi_l(t)|^2 - 1 ) \,,\\
  \eta_l(t) & = & g_l \chi_l^2(t)
\end{eqnarray}
that satisfy 
$\overline{\xi_l(t)}=\overline{\eta_l(t)}=\overline{\eta_l(t)\eta_{l'}(t')}=0$ 
as well as
\begin{eqnarray}
  \overline{\xi_l(t) \xi_{l'}(t')} & = & g_l g_{l'} J_{l'-l}^2
  \left(E_\delta (t'-t)/\hbar\right) \,,\\
  \overline{\eta_l^*(t)\eta_{l'}(t')} & = & (-1)^{l'-l}
  e^{-2 i E_\delta (t'-t) / \hbar} \overline{\xi_l(t) \xi_{l'}(t')} \,.
\end{eqnarray}
Without entering into too much technical detail, we can observe that
we are dealing here with continuous-time stochastic processes with zero
mean value whose amplitudes scale as $g_l/\hbar$ and which vary on a 
characteristic time scale $\hbar / E_\delta$.
This yields effective diffusion constants that scale as
$(g_l/\hbar)^2 (\hbar / E_\delta) = m g^2(l\delta) / \hbar^3$
and are therefore finite in the continuous limit $\delta\to 0$.

Note that the replacement $|\psi_l|^2 \to |\psi_l|^2 - 1$
in the translation of the nonlinear term from the Gross-Pitaevskii to the
tW equation is crucially important for the above argument to hold.
In colloquial terms, the effective potential $V^{\rm eff}_l = - g_l$ 
that is artificially introduced by this replacement exactly compensates 
for the enhanced repulsive interaction of the nonlinear wave with the 
vacuum fluctuations at positions where $g_l$ is particularly large.

\subsection{Numerical results}

\label{sec:results}

\begin{figure}
  \includegraphics[width=\columnwidth]{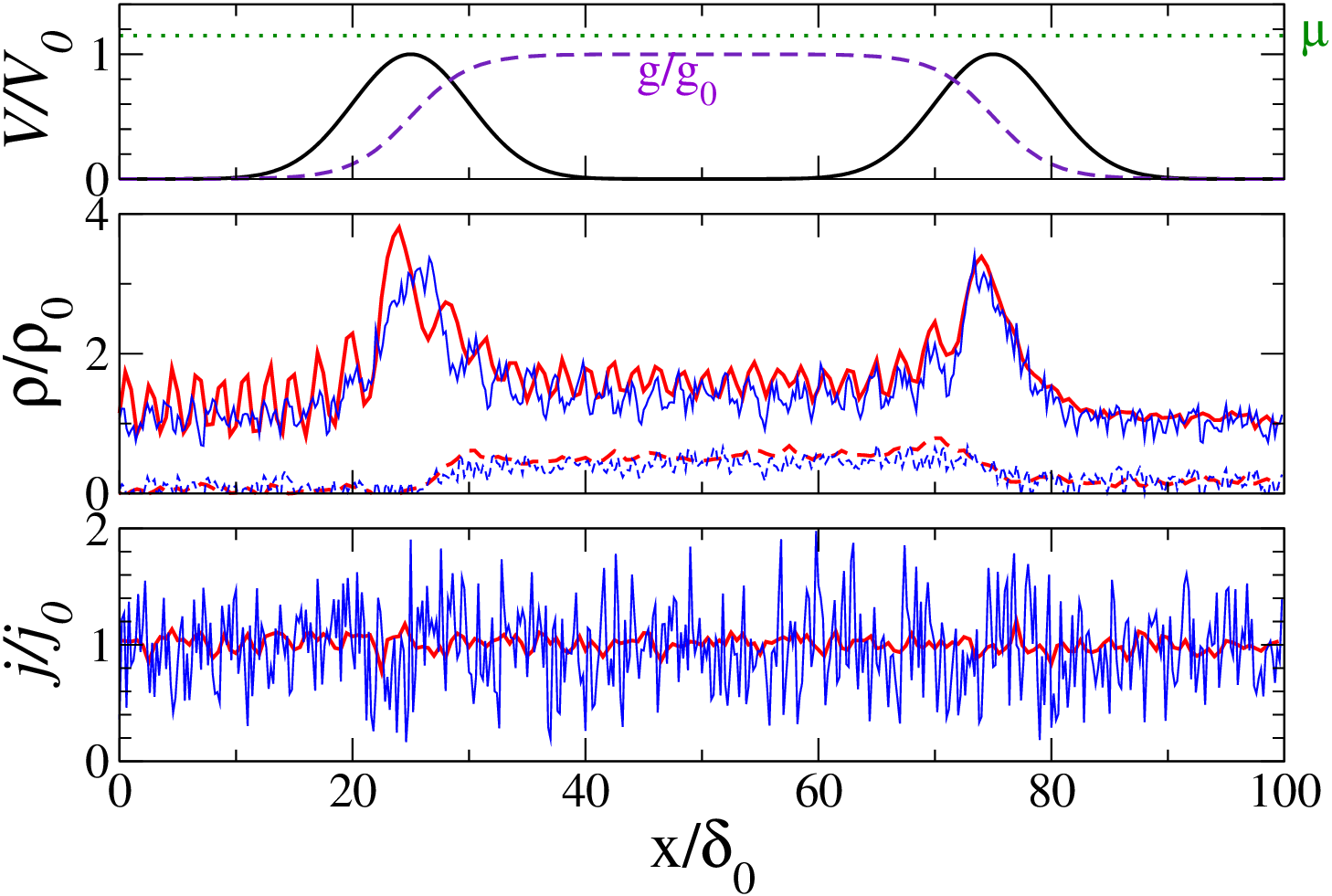}%
  \caption{\label{fig:prof1}
    Density and current profiles of the scattering state in the
    near-resonant transport configuration for two different grid spacings 
    $\delta = 0.5 \delta_0$ (thick red lines) and $\delta = 0.2 \delta_0$ 
    (thin blue lines).
    The upper panel shows the external potential $V(x)$ (solid black line)
    as well as the interaction strength $g(x)$ (violet dashed line)
    according to Eqs.~\eqref{eq:V} and \eqref{eq:g}, respectively.
    The green dotted line marks the level of the chemical potential.
    The middle panel shows the total densities (upper solid lines)
    and the incoherent parts of the density according to Eq.~\eqref{eq:dinc}
    (lower dashed lines), while the lower panel shows the total currents,
    respectively for $\delta = 0.5 \delta_0$ (thick red lines) and 
    $\delta = 0.2 \delta_0$ (thin blue lines).
    These tW calculations were done using $10^6$ random
    realizations of the vacuum fluctuations.}
\end{figure}

\begin{figure}
  \includegraphics[width=\columnwidth]{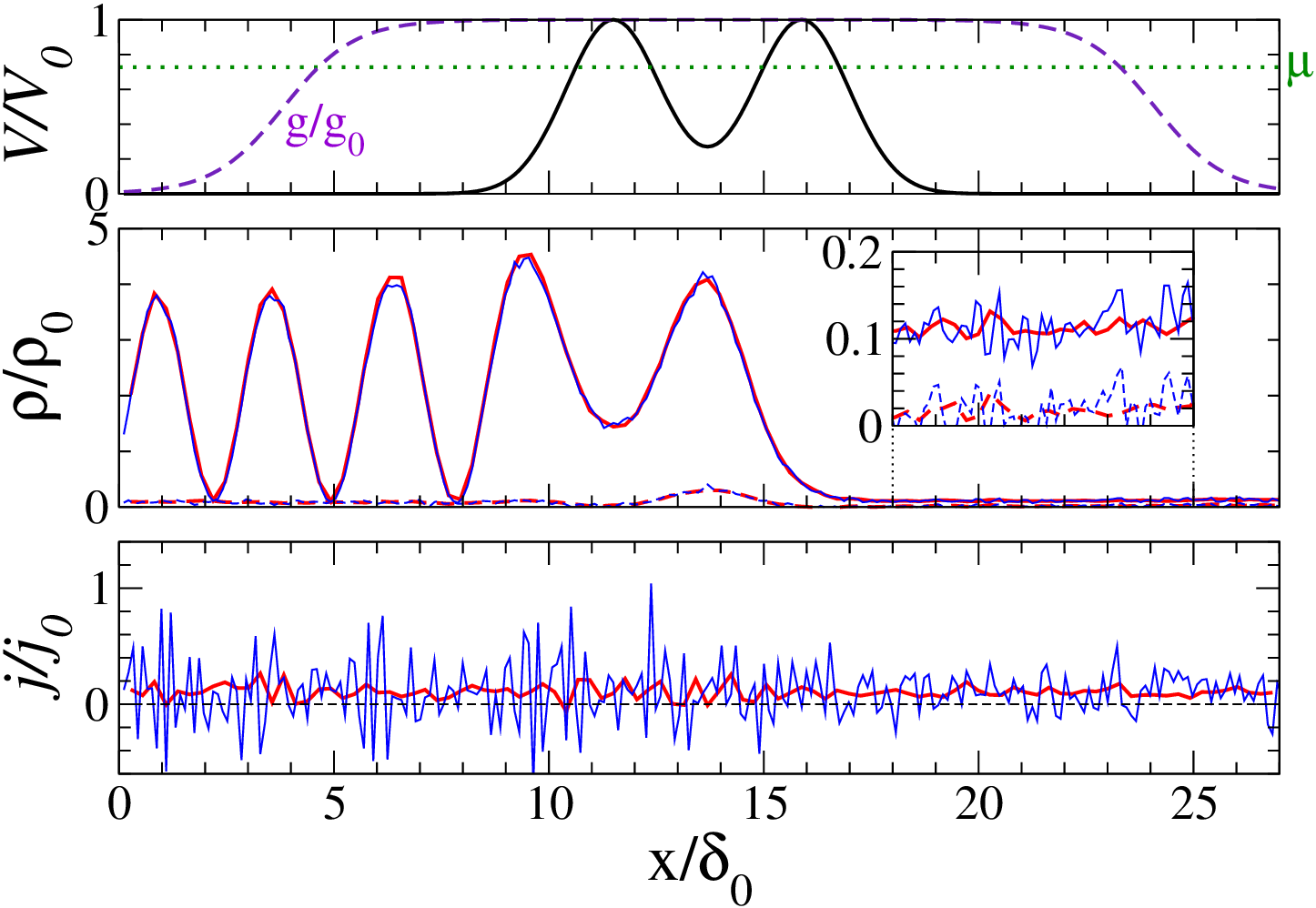}%
  \caption{\label{fig:prof2}
    Same as Fig.~\ref{fig:prof1} for the non-resonant transport configuration.
    Total and incoherent densities (middle panel) as well as total currents
    (lower panel) were computed for the grid spacings
    $\delta = 0.2739 \delta_0$ (thick red lines) and 
    $\delta = 0.1095 \delta_0$ (thin blue lines), using again 
    $10^6$ random realizations of the vacuum fluctuations.}
\end{figure}

Let us now verify these findings for the specific case of the transmission 
of a Bose-Einstein condensate through a symmetric double barrier potential.
The latter is formed by a sequence of two Gaussian barriers, i.e.
\begin{equation}
  V(x) = V_0 \left( e^{-(x-x_1)^2/(2\sigma^2)} + e^{-(x-x_2)^2/(2\sigma^2)} \right) \,,
  \label{eq:V}
\end{equation}
which are placed at the positions $x_1$ and $x_2$
and which have the same width $\sigma$.
The spatial dependence of the one-dimensional interaction strength 
is chosen as
\begin{equation}
  g(x) = g_0 \left( \tanh\left[\frac{x-\tilde{x}_1}{\tilde{\sigma}}\right] - 
  \tanh\left[\frac{x-\tilde{x}_2}{\tilde{\sigma}}\right] \right) \,.
  \label{eq:g}
\end{equation}
For the sake of simplicity, we consider the coupling strength to the trap 
to be strongly localized in space, such that we can write
$\sqrt{N}\kappa(x) = S_0 \delta(x)$.
This source emits a matter-wave beam whose chemical potential $\mu$ 
is tuned such that it is of the same order as the potential barrier height.
In the absence of the scattering potential and the interaction,
such a source would yield the homogeneous densities and currents
\begin{eqnarray}
  \rho_0 & = & |S_0|^2 m / ( 2 \hbar^2 \mu) \,, \\ 
  j_0 & = & \sqrt{2\mu/m} \rho_0 \,. 
\end{eqnarray}

We shall, in the following, consider two distinct transport scenarios:
a near-resonant one with well-separated potential barriers, leading to 
nearly perfect transmission of the matter-wave beam, and a non-resonant one
with overlapping barriers, leading to almost total reflection of the atoms.
The former (near-resonant) configuration is obtained by choosing the
parameters $x_1 = \tilde{x}_1 = 25 \delta_0$, 
$x_2 = \tilde{x}_2 = 75 \delta_0$, $\sigma = \tilde{\sigma} = 5 \delta_0$,
$g_0 = 0.034 V_0 \delta_0$, $S_0 = 0.8 V_0 \sqrt{\delta_0}$, and
$\mu = 1.15 V_0$. 
Here we introduce by $\delta_0 = \hbar/\sqrt{m V_0}$ the characteristic 
length scale that corresponds to the barrier height $V_0$.
For the latter (non-resonant) configuration, we choose
$x_1 = 11.50 \delta_0$, $x_2 = 15.88 \delta_0$, $\sigma = 1.095 \delta_0$,
$\tilde{x}_1 = 3.834 \delta_0$, $\tilde{x}_2 = 24.1 \delta_0$, 
$\tilde{\sigma} = 1.643 \delta_0$, $g_0 = 0.01826 V_0 \delta_0$,
$S_0 = 1.974 V_0 \sqrt{\delta_0}$, and $\mu = 0.727 V_0$.

The practical numerical implementation of the tW method in this waveguide 
scattering problem is done in the same way as described in 
Ref.~\cite{DujArgSch15PRA}.
In particular, we restrict the numerical representation of the waveguide
to a finite spatial region ($0<x<100\delta_0$ in the near-resonant and
$0<x<27.39\delta_0$ in the non-resonant case) in which the scattering 
potential and the interaction strength can be considered to be nonzero.
Within the framework of the tW approach, the coupling to 
the two noninteracting ``leads'' on the left- and right-hand side gives 
then rise to a time-dependent quantum noise that enters into the 
scattering region \cite{DujArgSch15PRA}.
The decay of atoms to the leads, on the other hand, is modeled through 
smooth exterior complex scaling \cite{DujSaeSch14APB}.

In the numerical practice, the source amplitude is slowly increased
with time from zero to its maximal value $S_0$.
Figure \ref{fig:prof1} shows the stationary density and current profiles that 
are numerically obtained after this ramping process in the near-resonant
transport configuration, for two different choices for the numerical grid 
spacing $\delta$.
We clearly see a nearly symmetric density profile, which is a signature
of near-resonant transmission.
The density maxima at the positions of the potential barriers as well as
the slight enhancement of the density in the interacting region in between
the barriers are explained by an effective decrease of the speed of the 
atoms due to energy conservation, which renders the atoms more likely to
be detected there.
As is indeed expected to occur in the presence of quasi-stationary scattering,
the current profile, on the other hand, is fairly homogeneous within and
outside the interacting region, apart from statistical fluctuations that
arise from a finite Monte-Carlo sampling in the framework of the tW method.

Figure \ref{fig:prof2} shows the stationary density and current profiles
in the non-resonant transport configuration, again for two different choices 
for the numerical grid spacing $\delta$.
While the chosen chemical potential $\mu = 0.727 V_0$ lies rather close to
a single-particle resonance of the double barrier potential, the presence of 
the atom-atom interaction gives rise to a significant shift of the effective 
resonance level to higher values of the chemical potential and thereby 
induces blocking of resonant transmission \cite{PauRicSch05PRL}.
Again, a fairly homogeneous current profile is obtained within and outside 
the interacting region.

\begin{figure}
  \includegraphics[width=\columnwidth]{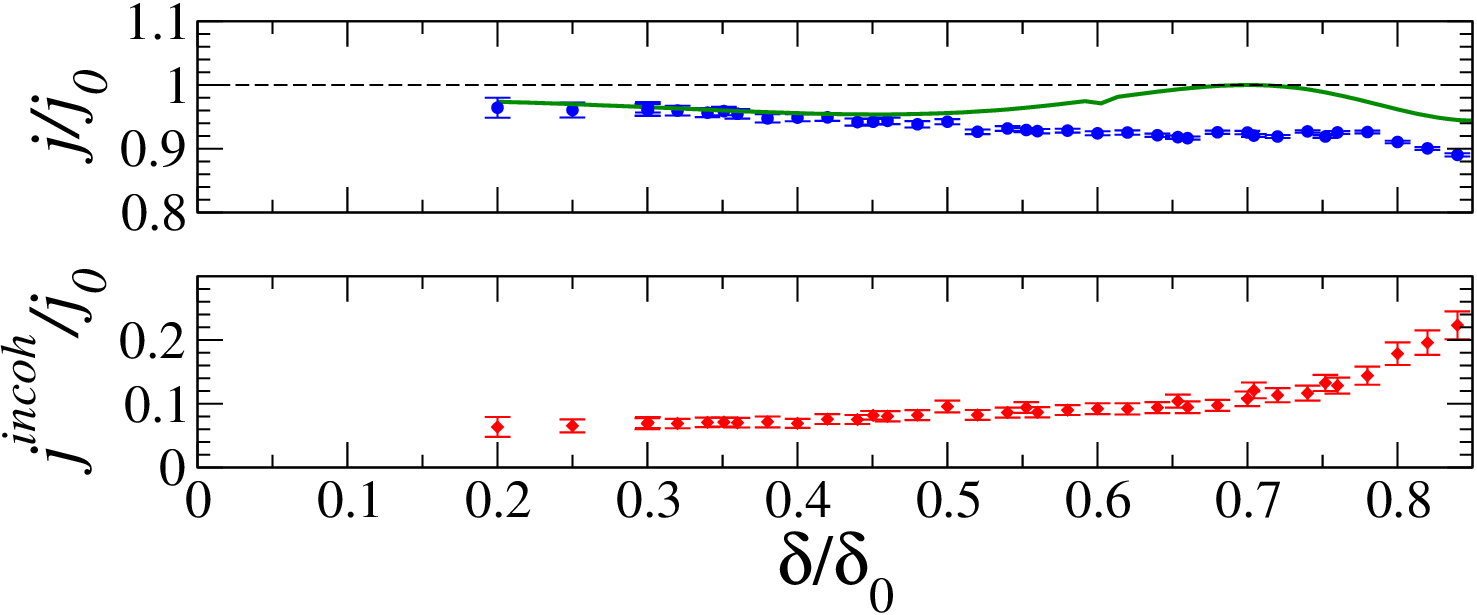}\\
  \includegraphics[width=\columnwidth]{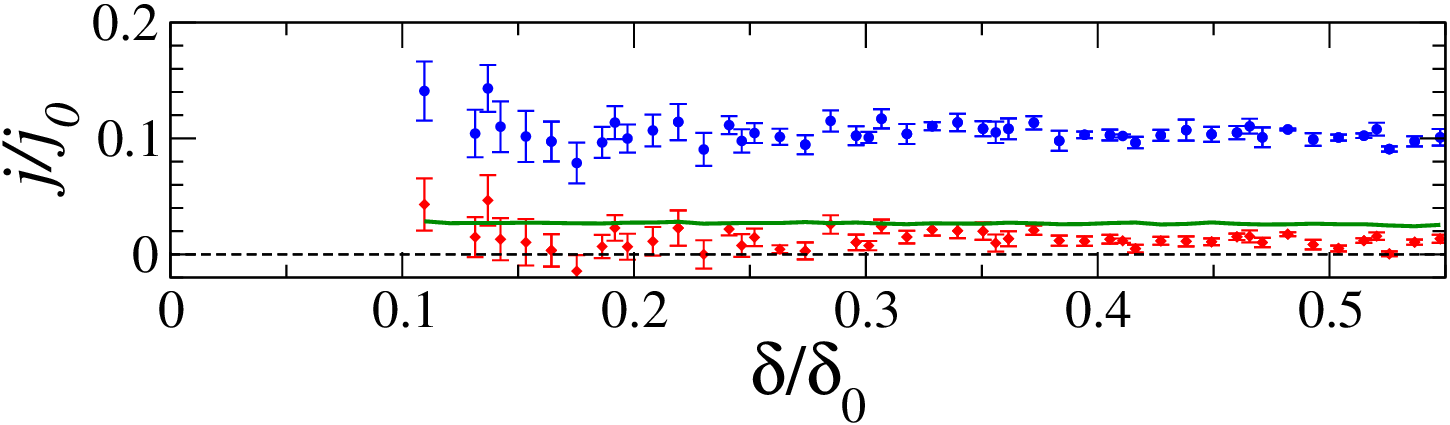}
  \caption{\label{fig:trans}
      Total (blue circles) and incoherent transmissions (red diamonds) 
      across the double barrier potential \eqref{eq:V}, obtained with the 
      tW method for different values of the numerical grid spacing $\delta$ 
      in the near-resonant (upper and middle panels) and the non-resonant
      transport configuration (lower panel) using the same parameters 
      of the scattering system as for Figs.~\ref{fig:prof1} and 
      \ref{fig:prof2}, respectively.
      The green solid lines show the results of a purely classical 
      Gross-Pitaevskii calculation according to Ref.~\cite{PauRicSch05PRL}.
      The total and incoherent currents are calculated according to 
      Eqs.~\eqref{eq:j} and \eqref{eq:jinc}, respectively, where we perform 
      an additional average over the numerical grid points of the system 
      (which then defines the depicted error bars).
      In both transport configurations, the total and the 
      incoherent transmission tend to finite values in the limit of 
      vanishing grid spacing $\delta\to 0$ \cite{rem_effort}.
  }
\end{figure}

Figure \ref{fig:trans} shows how the total and incoherent transmissions
across the double barrier potential scale with the grid spacing $\delta$.
We calculated for this purpose the average total current $j$ and the
incoherent current $j^{\rm incoh}$ according to Eqs.~\eqref{eq:j} and 
\eqref{eq:jinc}, respectively, and performed an additional average over 
the numerical grid points of the system, in order to reduce the size
of the statistical fluctuations.
Clearly, we see that both the total and the incoherent transmission 
tend to finite values in the limit of vanishing grid spacing $\delta\to 0$,
both for the near-resonant and the non-resonant transport configuration.
This numerical observation confirms that the tW method
is expected to yield consistent results in the continuous limit.

\section{Conclusion}

\label{sec:conc}

In summary, we investigated the feasibility of the tW approach for the 
description of quasi-statio\-nary scattering processes with interacting 
Bose-Einstein condensates in one-dimensional waveguides.
While a discretization of position space is most conveniently introduced
in order to apply the method in practice, we showed that consistent 
results are to be obtained in the continuous limit of vanishing grid spacing.
This finding provides promising perspectives for the applicability of the
tW method in the context of bosonic waveguide scattering processes.
Indeed, judging from Ref.~\cite{DujArgSch15PRA} we expect that the tW method
will yield reliable predictions in the presence of weak interaction strengths, 
similarly as for the Hartree-Fock-Bogoliubov method \cite{ErnPauSch10PRA}
or the Bogoliubov back-reaction method \cite{VarAng01PRL,TikAngVar07PRA}.
It certainly breaks down in the presence of very strong interactions where 
genuinely quantum many-body approaches based on the density matrix 
renormalization group (DMRG) and the matrix-product state (MPS) method 
\cite{Whi92PRL,Vid03PRL,Vid04PRL,VerPorCir04PRL,DalO04JSM} as well as on the 
Gutzwiller ansatz (see, e.g., Ref.~\cite{Zak05PRA}) are more appropriate 
to describe the dynamics of the system.

We furthermore presented a semiclassical derivation of the tW approach 
in the framework of the van Vleck-Gutzwiller theory, which essentially 
relates this approach to the diagonal semiclassical approximation.
This latter framework opens possibilities for a wider application of the tW
method in the context of moderately interacting systems that exhibit 
chaotic classical dynamics.
Indeed, while the conventional implementation of the tW method is not expected 
to yield reliable predictions for long evolution times that exceed the 
Ehrenfest time of such a chaotic system, \emph{average} transport observables, 
which are obtained, e.g., in the presence of disorder, are nevertheless 
expected to be correctly reproduced by this approach provided we can safely 
assume classical ergodicity (and account for systematic quantum interference 
effects such as coherent backscattering in Fock space \cite{EngO14PRL}).
This new semiclassical perspective of the tW approach will also form a useful 
basis for the development of a truly semiclassical van Vleck-Gutzwiller 
(or Herman-Kluk) approach that is able to capture interference effects beyond 
the diagonal approximation \cite{EngO14PRL,SimStr14PRA,EngUrbRic14xxx}.

\begin{acknowledgements}
This work was financially supported by the DFG Research Unit FOR760 
as well as by a ULg research and mobility grant for T.E.
Computational resources have been provided by the 
\textit{Consortium des \'{E}quipements de Calcul Intensif (C\'{E}CI)}, 
funded by the F.R.S.-FNRS under Grant No. 2.5020.11.
\end{acknowledgements}


\begin{thebibliography}{[10]}
\bibitem{rem_exact}
For the exact quantum mechanical equivalent of Eq.~\eqref{eq:AtW}, $A[{\bold
  q}^{\rm f},{\bold p}^{\rm f}]$ would have to be replaced by the Wigner
  transform of the associated time-dependent operator $\hat{A}$ which is
  unavailable in practice.
\bibitem{rem_effort}
Note that the numerical effort that is necessary to compute the tW averages of
  the total and incoherent currents with good statistical convergence increases
  dramatically with decreasing grid spacing. We therefore do not show
  transmission data for grid spacings $\delta < 0.2 \delta_0$ in the
  near-resonant and $\delta < 0.1 \delta_0$ in the non-resonant case.
\end{thebibliography}

\providecommand{\WileyBibTextsc}{}
\let\textsc\WileyBibTextsc
\providecommand{\othercit}{}
\providecommand{\jr}[1]{#1}
\providecommand{\etal}{~et~al.}

\end{document}